\documentclass[sigplan,10pt,nonacm]{acmart} 
\renewcommand\footnotetextcopyrightpermission[1]{} % removes footnote with conference information in first column
\setcopyright{none}
\usepackage{booktabs}   %% For formal tables:
\usepackage{subcaption} %% For complex figures with subfigures/subcaptions
\usepackage[utf8]{inputenc}
\inputencoding{utf8}
\usepackage[T1]{fontenc}
\usepackage{amsmath}
\usepackage{amsfonts}
\usepackage{amssymb}
\usepackage[english]{babel}
\usepackage[nounderscore]{syntax}
\usepackage[inference]{semantic}

\usepackage{parskip}
\usepackage{url}
\usepackage{listings}
\usepackage{color}
\usepackage{xcolor}
\usepackage{titlesec}
\usepackage{textcomp}
\usepackage{booktabs} % For formal tables

\usepackage{blindtext}
\usepackage{enumitem}
\usepackage{flushend}
\usepackage{microtype}
\usepackage{hyperref}
\usepackage{float}
\usepackage{placeins}

\definecolor{mygreen}{rgb}{0,0.6,0}
\definecolor{mygray}{rgb}{0.5,0.5,0.5}
\definecolor{mymauve}{rgb}{0.58,0,0.82}

\makeatletter
\lstdefinelanguage{llvm}{
  morecomment = [l]{;},
  morestring=[b]", 
  sensitive = true,
  classoffset=0,
  morekeywords={
    define, declare, global, constant,
    internal, external, private,
    linkonce, linkonce_odr, weak, weak_odr, appending,
    common, extern_weak,
    thread_local, dllimport, dllexport,
    hidden, protected, default,
    except, deplibs,
    volatile, fastcc, coldcc, cc, ccc,
    x86_stdcallcc, x86_fastcallcc,
    ptx_kernel, ptx_device,
    signext, zeroext, inreg, sret, nounwind, noreturn,
    nocapture, byval, nest, readnone, readonly, noalias, uwtable,
    inlinehint, noinline, alwaysinline, optsize, ssp, sspreq,
    noredzone, noimplicitfloat, naked, alignstack, inaccessiblememonly, speculatable,
    module, asm, align, tail, to,
    addrspace, section, alias, sideeffect, c, gc,
    target, datalayout, triple,
    blockaddress
  },
  classoffset=1, keywordstyle=\color{purple},
  morekeywords={
    fadd, sub, fsub, mul, fmul,
    sdiv, udiv, fdiv, srem, urem, frem,
    and, or, xor,
    icmp, fcmp,
    eq, ne, ugt, uge, ult, ule, sgt, sge, slt, sle,
    oeq, ogt, oge, olt, ole, one, ord, ueq, ugt, uge,
    ult, ule, une, uno,
    nuw, nsw, exact, inbounds,
    phi, call, select, shl, lshr, ashr, va_arg,
    trunc, zext, sext,
    fptrunc, fpext, fptoui, fptosi, uitofp, sitofp,
    ptrtoint, inttoptr, bitcast,
    ret, br, indirectbr, switch, invoke, unwind, unreachable,
    malloc, alloca, free, load, store, getelementptr,
    extractelement, insertelement, shufflevector,
    extractvalue, insertvalue,
  },
  alsoletter={\%},
  keywordsprefix={\%},
}
\makeatother

\usepackage{appendix}
\usepackage{booktabs}
\usepackage{hyperref}
\usepackage{lmodern}
\usepackage{multirow}
\usepackage{graphicx}
\graphicspath{ {./assets/} }
\usepackage{xspace}

\newcommand{\invariantgroup}{\lstinline{invariant.group}\xspace}
\newcommand{\launder}{\lstinline{launder}\xspace}
\newcommand{\strip}{\lstinline{strip}\xspace}

\begin{document}

%% Title information
\title{Modeling the Invariance of Virtual Pointers in LLVM}         %% [Short Title] is optional;
                                        %% when present, will be used in
                                        %% header instead of Full Title.
%\titlenote{with title note}             %% \titlenote is optional;
                                        %% can be repeated if necessary;
                                        %% contents suppressed with 'anonymous'
%\subtitle{Subtitle}                     %% \subtitle is optional
%\subtitlenote{with subtitle note}       %% \subtitlenote is optional;
                                        %% can be repeated if necessary;
                                        %% contents suppressed with 'anonymous'

%% Author information
%% Contents and number of authors suppressed with 'anonymous'.
%% Each author should be introduced by \author, followed by
%% \authornote (optional), \orcid (optional), \affiliation, and
%% \email.
%% An author may have multiple affiliations and/or emails; repeat the
%% appropriate command.
%% Many elements are not rendered, but should be provided for metadata
%% extraction tools.

%% Author with single affiliation.
\author{Piotr Padlewski}
%\authornote{with author1 note}          %% \authornote is optional;
                                        %% can be repeated if necessary
\orcid{0000-0001-5107-0824}             %% \orcid is optional
\affiliation{
  %\position{Position1}
  %\department{Department1}              %% \department is recommended
  \institution{Google Research}            %% \institution is required
  %\streetaddress{Street1 Address1}
  %\city{City1}
  %\state{State1}
  %\postcode{Post-Code1}
  %\country{Country1}                    %% \country is recommended
}
\email{prazek@google.com}          %% \email is recommended

\author{Krzysztof Pszeniczny}
%\authornote{with author1 note}          %% \authornote is optional;
                                        %% can be repeated if necessary
\orcid{0000-0002-9115-0893}             %% \orcid is optional
\affiliation{
  %\position{Position1}
  %\department{Department1}              %% \department is recommended
  \institution{Google Research}            %% \institution is required
  %\streetaddress{Street1 Address1}
  %\city{City1}
  %\state{State1}
  %\postcode{Post-Code1}
  %\country{Country1}                    %% \country is recommended
}
\email{kpszeniczny@google.com}          %% \email is recommended

\author{Richard Smith}
%\authornote{with author1 note}          %% \authornote is optional;
                                        %% can be repeated if necessary
%\orcid{nnnn-nnnn-nnnn-nnnn}             %% \orcid is optional
\affiliation{
  %\position{Position1}
  %\department{Department1}              %% \department is recommended
  \institution{Google Research}            %% \institution is required
  %\streetaddress{Street1 Address1}
  %\city{City1}
  %\state{State1}
  %\postcode{Post-Code1}
  %\country{Country1}                    %% \country is recommended
}
\email{richardsmith@google.com}          %% \email is recommended
% TODO: do you have orcid?

%% Abstract
%% Note: \begin{abstract}...\end{abstract} environment must come
%% before \maketitle command
\begin{abstract}
Devirtualization is a compiler optimization that replaces indirect (virtual) function calls with direct calls.  It is particularly effective in object-oriented languages, such as Java or C++, in which virtual methods are typically abundant.

We present a novel abstract model to express the lifetimes of C++ dynamic objects and invariance of virtual table pointers in the LLVM intermediate representation. The model and the corresponding implementation in Clang and LLVM enable full devirtualization of virtual calls whenever the dynamic type is statically known and elimination of redundant virtual table loads in other cases.

Due to the complexity of C++, this has not been achieved by any other C++ compiler so far.  Although our model was designed for C++, it is also applicable to other languages that use virtual dispatch.  Our benchmarks show an average of 0.8\% performance improvement on real-world C++ programs, with more than 30\% speedup in some cases.  The implementation is already a part of the upstream LLVM/Clang and can be enabled with the \lstinline{-fstrict-vtable-pointers} flag.
\end{abstract}

%% 2012 ACM Computing Classification System (CSS) concepts
%% Generate at 'http://dl.acm.org/ccs/ccs.cfm'.
\begin{CCSXML}
<ccs2012>
<concept>
<concept_id>10011007.10011006.10011041</concept_id>
<concept_desc>Software and its engineering~Compilers</concept_desc>
<concept_significance>100</concept_significance>
</concept>
</ccs2012>
\end{CCSXML}

\ccsdesc[100]{Software and its engineering~Compilers}
%% End of generated code

%% Keywords
%% comma separated list
\keywords{devirtualization, invariant, virtual pointer, indirect call, fat pointer} %,vptr,indirect call,fat pointer,-fstrict-vtable-pointers}  %% \keywords are mandatory in final camera-ready submission

\hypersetup{
pdftitle={Modeling the Invariance of Virtual Pointers in LLVM},
pdfsubject={Computer Science; Compilers},
pdfauthor={Piotr Padlewski, Krzysztof Pszeniczny, Richard Smith},  
pdfkeywords={Devirtualization; virtual call; virtual pointer; vptr; indirect call; virtual table; polymorphism; C++; LLVM; Clang}
}

%% \maketitle
%% Note: \maketitle command must come after title commands, author
%% commands, abstract environment, Computing Classification System
%% environment and commands, and keywords command.
\maketitle

\section{Introduction}

\inputencoding{utf8}

Devirtualization is a compiler
optimization that changes virtual (dynamic) calls to direct (static) calls.
The former introduce a performance penalty compared to the latter, as they cannot be inlined and are harder to speculatively execute. 

Indirect calls can also serve as an attack vector if an adversary can replace the virtual pointer of an object, e.g. by performing buffer overflow.

Moreover, a recently discovered vulnerability called Spectre~\cite{Spectre,Kocher2018spectre} showed that the indirect branch predictor is susceptible to side-channel attacks where observable side effects could lead to private data leakage.
The currently used mitigation technique (the so-called `retpolines'~\cite{Retpolines}) effectively eliminates indirect branch prediction entirely.

Indirect calls also limit the extent of compiler analyses and optimizations, making it much more difficult to perform inter-procedural reasoning and thus hinder many useful intra-procedural optimizations, e.g. inlining, intra-procedural register allocation, constant propagation or function attributes inference.

Modeling the lifetime of a C++ pointer in the LLVM IR is nontrivial, because bit-wise equal C++ pointers, viz. having the same bit pattern, may nevertheless denote objects with different lifetimes. They are thus not equivalent and cannot be replaced with each other. However, this information is not reflected in the LLVM IR, because equal values are interchangeable there. The following C++ code, when translated to the LLVM IR and optimized\footnote{The LLVM IR code was simplified to make it clearer. Mangled names, types or unimportant attributes will be similarly simplified in other listings.}, makes the issue apparent:

\begin{lstlisting}[language=c++]
void test() {
    auto* a = new A;
    external_fun(a);

    A* b = new(a) B;
    external_fun(b);
}
\end{lstlisting}

\begin{lstlisting}[language=llvm]
define void @test() {
    %new = call i8* @NEW(i64 8)
    %a = bitcast i8* %1 to %struct.A*
    call void @A_CTOR(%struct.A* %a)
    call void @external_fun(%struct.A* %a)
    %b = bitcast i8* %1 to %struct.B*
    call void @B_CTOR(%struct.B* %b)
    ; Because %a and %b are known to be equal,
    ; the compiler decided to use %a.
    call void @external_fun(%struct.A* %a)
    ret void
}
\end{lstlisting}

Here, the so-called ``placement new'' was used to construct a new object in pre-allocated memory (here: in the memory pointed by \lstinline{a}).
As can be seen in this example, \lstinline{%a} is passed to the second call of \lstinline{external_fun} instead of \lstinline{%b}, as the compiler figured out that they are equivalent. On the other hand, if \lstinline{external_fun(b)} had been replaced with \lstinline{external_fun(a)}, the behavior would be undefined per the C++ standard.

In this paper, we present a novel way to model C++ dynamic objects' lifetimes in LLVM capable of leveraging their virtual pointer's invariance to aid devirtualization. This is a previously unresolved problem, blocking the wide deployment of devirtualization techniques.

\section{Related Work}
There are multiple optimizations done in hardware that speed up the execution of virtual calls. The most important one is the indirect branch predictor that makes speculative execution of indirect calls possible. Moreover, some CPU architectures provide store buffers~\cite[11.10 Store Buffer]{IntelManual}
~\cite[2.4.5.2 L1 DCache]{IntelOptimizationManual}~\cite[8.5.1. Store buffer]{StoreBufferArm}~\cite{Owens:2009:BXM:1616077.1616107}, which temporarily hold writes before committing them to memory. When a load operation is performed, the CPU returns a value from the store buffer if it exists for the given address. Thus, after construction of an object, as long as its virtual pointer's value is in the store buffer, loading it is more efficient.

Additionally, there are also multiple software techniques used for devirtualization. Speculative devirtualization~\cite{HubickaDevirtualization3,HubickaDevirtualization4} introduces direct calls to known implementations by guarding them with comparisons of virtual pointers or virtual function pointers. This enables inlining and relies on the branch predictor instead of the indirect call predictor, which is more accurate especially on older architectures. However, only implementations visible in the current translation unit can be speculated. An indirect call is thus still required for unknown implementations.

A further similar technique called indirect call promotion~\cite{ICP} uses profiling data of the program to pick the most frequent callsites and insert direct calls guarded by comparison of function pointers similarly to speculative devirtualization.

Another approach is to use information about the entire program using a family techniques called whole program optimizations (WPO), also referred as link-time optimization (LTO)~\cite{LTO,LTO_GCC,Johnson2017ThinLTOSA,ThinLTOBlog,WHOPR} as it often happens during link-time.
Having information about the whole program, a compiler can derive facts that might be unknown or hard to model in a single translation unit, e.g. a virtual pointer being invariant, or the definition of virtual tables, or that a function is effectively final (meaning it is not overridden by other types).
Using information about the whole program, GCC in LTO mode can prune unreachable branches created by speculative devirtualization. LLVM's whole-program devirtualization~\cite{WPD} is also able to identify virtual functions that only return constants, and then put these constants in the vtable itself and replace calls with simple loads.

Similar optimizations can be also achieved using just-in-time compilation. However, it is rarely used in the case of C++.

Another technique is to leverage the invariance of a virtual pointer and having loaded the pointer once, reuse it for multiple virtual calls, or even devirtualize all virtual calls if the dynamic type of object is statically known. The latter holds e.g. for automatic objects in C++.

In Java, the virtual pointer is set only once in the most derived constructor.  There is also no mechanism for changing the dynamic type of an existing object, and the lack of manual memory management means that no reference ever points to a dead object. Hence, one can simply inform the optimizer that virtual pointers are invariant. This technique was used in Falcon~\cite{Falcon, Falcon2} -- an LLVM-based Java JIT compiler, resulting in 10-20\% speedups\footnote{Personal communication from Philip Reames, Director of Compiler Technology at Azul Systems}.

Moreover, some LLVM front-ends decided to create a richer language-specific intermediate language used between abstract syntax tree and the LLVM IR, such as SIL (Swift), MIR (Rust), or Julia IR (Julia). Unfortunately, Clang translates its AST directly to the LLVM IR.

Previous experimental LLVM-based models~\cite{DevirtualizationInLLVM, Padlewski:2017:DL:3135932.3135947, DevirtualizationBlogpost, RFCDevirtualization} that tried to express lifetime of pointers to dynamic objects were unfortunately flawed. They failed to prevent the the compiler from being easily confused by equality of pointers to objects with different lifetimes. Hence, they could not be enabled by default, as they could lead to a miscompilation, e.g.:

\begin{minipage}{\linewidth}
\begin{lstlisting}[language=C++]
void g() {
    A *a = new A;
    a->virt_meth();
    A *b = new(a) B;
    if (a == b) {
        // Here the compiler is exposed to the fact
        // that a == b, so it may replace the SSA
        // value of b with a, which would result in
        // an erroneous call to A::virt_meth.
        b->virt_meth();
    }
}
\end{lstlisting}
\end{minipage}

\section{Problems Specific To C++}
\label{sec:C++}
In C++ functions are allowed to modify any accessible memory, so the compiler needs to be very conservative with any assumptions.
Moreover, compilation of separate modules -- called
translation units~\cite[5.2 Phases of translation]{CPP} -- is usually independent, which makes compilation highly parallelizable, but on the other hand limits the ability of the compiler to reason about functions from other translation units.

External functions are functions defined in a different translation unit than the currently compiled one, so that the body of a function is not visible to the compiler, unless performing whole program optimization. External functions are problematic when reasoning about virtual calls in C++ because without additional knowledge the compiler has to conservatively assume that the virtual pointer might be clobbered (overwritten).

The following code snippet demonstrates some of the shortcomings of traditional C++ devirtualization techniques that do not rely on any additional notion of virtual pointer invariance. 

\begin{minipage}{\linewidth}
\begin{lstlisting}[language=c++]
struct A {
    virtual void virt_meth();
};

void external_fun(A *a);
void foo(A *a) {
    auto *a = new A;
    // Conservatively assumed to potentially clobber
    // a's virtual pointer.
    external_fun(a);
    a->virt_meth();
}

void bar() {
    auto a = new A;
    a->virt_meth();
    // Can be devirtualized only if the first call
    // was inlined, as virt_meth is conservatively
    // assumed to potentially clobber a's virtual
    // pointer.
    a->virt_meth();
}
\end{lstlisting}
\end{minipage}

In C++ objects can change their dynamic type multiple times during construction and destruction, as each constructor/destructor sets the virtual pointer to the type of the class defining them. Moreover, placement new can create a new object in the memory previously occupied by a different one. This could even happen inside a virtual call, which is allowed to call placement new on the \lstinline{this} pointer:

\begin{minipage}{\linewidth}
\begin{lstlisting}[language=c++]
void A::virt_meth() {
    new(this) B;
}
\end{lstlisting}
\end{minipage}

Although calling placement new with \lstinline{this} is allowed, one would nevertheless be unable to use an old reference or pointer to such object afterwards, as they become invalid with the end of the lifetime of the pointed-to object~\cite[6.6.3
Object lifetime]{CPP}.
The only way to reuse these memory locations would be to use the pointer returned by placement new, or call \lstinline{std::launder}, which is the standardized way of obtaining a usable pointer located at the passed address.

Consider the following C++ code:

\begin{minipage}{\linewidth}
\begin{lstlisting}[language=c++]
struct A {
    int value;
};

struct B : A { };

void test() {
    auto *a = new A;
    a->value = 0;
    A *b = new(a)B;
    b->value = 42;
    if (a == b) {
        printf("%d %d", a->value, b->value);
    }
}
\end{lstlisting}
\end{minipage}

The behavior of this piece of code is undefined, because although the pointers \lstinline{a} and \lstinline{b} refer to the same memory location and are bit-wise equal, \lstinline{a} cannot be dereferenced after calling placement new, as the latter ends the object's lifetime. Note that reading the value of a pointer to a dead object is legal, but dereferencing it is not. The latter, however, can be safely done by using \lstinline{std::launder}:

\begin{minipage}{\linewidth}
\begin{lstlisting}[language=c++]
if (a == b) {
    // Fine, `a` is not referenced here.
    printf("%d %d", std::launder(a)->value,
           b->value);
}
\end{lstlisting}
\end{minipage}

The C++ rules allow the compiler to assume that whenever multiple virtual calls are performed through the same pointer, they all refer to the same dynamic type.

\begin{minipage}{\linewidth}
\begin{lstlisting}[language=c++]
void multiple_calls(A *a) {
    a->virt_meth();
    // Because `a` is dereferenced again, one may
    // assume that the dynamic type of `*a` did not
    // change; otherwise the behavior would have
    // been undefined. Hence no vtable reload is
    // needed here, the same function is called.
    a->virt_meth();
}
\end{lstlisting}
\end{minipage}

As an interesting corner case, it is nevertheless allowed to dereference a pointer after changing its dynamic type, as long as the dynamic type was changed back to the original one:

\begin{minipage}{\linewidth}
\begin{lstlisting}[language=c++]
void A::virt_meth() {
    // Changing dynamic type of 'this'.
    auto *b = new(this) B;
    // Calling virtual functions on 'this' is now
    // not allowed. However, calling them on 'b' 
    // is fine.
    b->virt_meth();
    new(this) A;
    // Can now legally use 'this'.
    other_virt_meth();
}
\end{lstlisting}
\end{minipage}

We leverage the above semantics in order to model virtual pointers as invariant, solving the issues described in examples above, which no existing compilers can solve at the time of writing.

\inputencoding{utf8}

\section{The Model}
\label{ch:Model}
We propose an abstract model that allows us to reason about dynamic objects and their use, including object creation, performing virtual calls, and changing dynamic types of existing objects.
Such a description is crucial in ensuring correctness of a translation from C++ to the LLVM IR in a way that does not inhibit optimizations.

We model pointers to objects of dynamic type as ``fat pointers'' -- pointers that carry additional information alongside a memory location (address). In our case, we would like to think of them as pointers that also store the current dynamic type.
Virtual calls will load the dynamic information from a fat pointer instead of the virtual pointer.

Creation of a fat pointer is achieved by a call to an intrinsic (a built-in function) called \launder (as a reference to \lstinline{std::launder}), that for a given address
returns a fat pointer that is valid to use, as long as the given memory location is occupied by a live object.
We claim that calls to \lstinline{std::launder} should be translated to calls of our intrinsic, described below.

In our model, accessing a class member (field) or comparing pointers to dynamic objects requires stripping dynamic information from a fat pointer -- we call this operation \strip. Stripping is needed because those operations cannot rely on the dynamic type of object; for example, comparison between pointers should compare only the addresses and not the dynamic information.

Because a fat pointer contains information about the current dynamic type, reading the dynamic type is a pure operation that depends only on the value of the fat pointer. This means that we can model the pureness of this operation as a property of a load instruction.

\subsection{Properties of the Model}
\label{subsec:Properties of the Model}
\begin{itemize}
    \item \lstinline{strip} is a pure operation, i.e., its value depends only on
    its argument.
    \item \lstinline{launder} is \textbf{not} a pure operation: it creates a
    fresh fat pointer with the current dynamic type each time it is executed.
    \item \lstinline!strip(strip(X))! and \lstinline!strip(launder(X))! are both equal to \lstinline!strip(X)!.
    This is because the model does not care what dynamic information is stripped.
    \item \lstinline!launder(strip(X))! and \lstinline!launder(launder(X))! may
    be replaced with \lstinline{launder(X)}.
    Note that this does not mean that \lstinline{launder(launder(X)) == launder(X)} --
    launder is inherently nondeterministic and no two invocations of it ever
    return the same value.
    \item A pointer passed to \lstinline{launder} or \lstinline{strip} aliases (the same memory
    is accessible through different pointers) the returned pointer.
\end{itemize}

\section{Modeling in LLVM}
\label{sec:Modeling in LLVM}
In this section, we show how to translate our model to the LLVM IR. One way to represent the fat pointer is as a struct containing an address and some additional metadata about the dynamic type, e.g. the virtual pointer. Then, every virtual call can use additional information instead of loading the virtual pointer from the object.
This achieves its goals because the fat pointer does not change during the object's lifetime, as the same SSA value is used for every virtual call on the given object.
However, we observe that in such a case, the virtual pointer would be stored in two places -- as a member in class and as extra data in fat pointer.
Unfortunately, removing the virtual pointer from the class instances is unacceptable in a real-world C++ compiler as
it breaks the established Application Binary Interface (e.g Itanium ABI~\cite{ItaniumABI}) -- a specification
describing the common behavior between the binary files (object files, executable, or a library), like the layout of the objects or calling conventions. Thus, it would not be possible to link ordinary object code with object code that uses fat pointers.

Fortunately, we can remove the additional information from the fat pointer and model the fact that loads of the virtual pointer are invariant as a property of the load instruction.
This is beneficial, as it does not introduce any data redundancy and does not break the ABI.
To model the fact that dynamic information is invariant for any given fat pointer, we introduce new instruction metadata -- \lstinline{invariant.group} -- that can be attached to a \lstinline{load} or \lstinline{store} instruction.
Instruction metadata in LLVM hints to the optimizer about special properties of an instruction. As specified by the LLVM language reference, stripping instruction metadata should not change the semantics of the program.

The presence of the \lstinline{invariant.group} metadata on the instruction instructs the optimizer that every load and store to the same pointer operand can be assumed to load or store the same value.
We attach this metadata on vtable loads (when performing virtual calls) and stores (in constructors and destructors).

To model the fact that the vtable contents are constant and do not change during the execution of the program, we use \lstinline{invariant.load} metadata. This metadata specifies that the memory location referenced by the load contains the same value at all points in the program.  Note that this requires a stronger assumption than \invariantgroup. With \lstinline{invariant.load}, multiple loads of the same function pointer from one vtable may be merged into one value.

Modeling \lstinline{strip} and \lstinline{launder} is done by introducing new intrinsics that take and return regular pointers, carrying imaginary meta information.  The intrinsics only serve as annotations in the IR and are stripped just before the code generation phase, hence they do not introduce any runtime or object code size cost.

\subsection{llvm.strip.invariant.group}
\label{subsec:llvm.strip.invariant.group}

The \strip operation is represented by a new intrinsic function \lstinline[language=llvm]{i8* @llvm.strip.invariant.group(i8*)}, used when an invariant established by \invariantgroup metadata no longer holds, to obtain a new pointer value that does not carry the invariant information.
It has the \lstinline[language=llvm]{readnone}, \lstinline[language=llvm]{speculatable}, and \lstinline[language=llvm]{nounwind} attributes~\cite[Function Attributes]{LangRef} meaning that the function is pure, can be speculated\cite[Function Attributes]{LangRef} and that it does not throw exceptions.

\subsection{llvm.launder.invariant.group}
\label{subsec:llvm.launder.invariant.group}
The \launder operation is represented by a new intrinsic function \lstinline[language=llvm]{i8* @llvm.launder.invariant.group(i8*)}, used when an invariant established by \invariantgroup metadata no longer holds, to obtain a new pointer that carries fresh invariant group information.

It has \lstinline[language=llvm]{inaccessiblememonly} attribute meaning that this function can modify only memory that is not visible to the current module. In this context, we can think of it as storing the additional virtual pointer in some imaginary location. Similarly to \strip, it has the \lstinline[language=llvm]{speculatable} and \lstinline[language=llvm]{nounwind} attributes.

The following code sample illustrates the intrinsics:
\begin{lstlisting}[language=llvm]
%p = alloca i8
store i8 42, i8* %p, !invariant.group !{}
call void @foo(i8* %p)

; *%p is known not to have changed
%a = load i8, i8* %p, !invariant.group !{}
call void @foo(i8* %p)

%unknown = load i8, i8* @unknown_ptr

; All stores to *%p write equal values,
; hence we may infer that %unknown == 42
store i8 %unknown, i8* %p, !invariant.group !{}

call void @foo(i8* %p)
%q = call i8* @llvm.launder.invariant.group(i8* %p)
; Cannot propagate invariant.group information
; through an llvm.invariant.group.launder and
; thus %d cannot be replaced with %a
%d = load i8, i8* %q, !invariant.group !{}
\end{lstlisting}
Here loads from the pointer \lstinline{%p} can be assumed to load the same value when \invariantgroup is present, and the load of \lstinline{%q} returned by the \launder cannot.

One of the primary concerns when making the model sound was that comparison of bit-wise equal pointers, representing objects of different lifetimes, should not make it possible to make them equivalent, meaning that one pointer could be replaced with the second one.

For example, consider again the code snippet from the introduction to this paper, that demonstrated a flaw present in the previous models:

\begin{minipage}{\linewidth}
\begin{lstlisting}[language=C++]
void g() {
    A *a = new A;
    a->virt_meth();
    A *b = new(a) B;
    if (a == b) {
        // Here the compiler is exposed to fact
        // that a == b, so it may replace the SSA
        // value of b with a, which would result in
        // an erroneous call to A::virt_meth.
        b->virt_meth();
    }
}
\end{lstlisting}
\end{minipage}

Under our model, the dynamic information is always stripped whenever an operation that can expose it is used, e.g. pointers about to be compared are always passed through a \strip:

\begin{minipage}{\linewidth}
\begin{lstlisting}[language=llvm]
%vtable_a = load i8* %a, !invariant.group !{}
; ...
; if (a == b)
%sa = call i8* @llvm.strip.invariant.group(i8* %a)
%sb = call i8* @llvm.strip.invariant.group(i8* %b)
%bool = icmp %sa, %sb
br %bool, %if, %after
if:
; a == b was lowered to %sa == %sb which does not
; imply %a == %b, thus %vtable_b cannot be replaced
; with %vtable_a
%vtable_b = load i8* %b, !invariant.group !{}
\end{lstlisting}
\end{minipage}

\subsection{Emitting Launder and Strip}
\label{subsec:Emitting Launder And Strip}
We apply the following rules to decide when to emit \launder or \strip when translating C++ to the LLVM IR:
\begin{itemize}
    \item Constructors of derived classes need to \launder the \lstinline{this} pointer
before passing it to the constructors of base classes; they may subsequently
operate on the original \lstinline{this} pointer.
    \item Likewise, destructors of derived classes must \launder the \lstinline{this}
pointer before passing it to the destructors of base classes.
    \item Placement new and \lstinline{std::launder} shall call \launder.
\item Accesses to union members must call \launder, because the active member
of the union may have changed since the last visible access.
\item Whenever two pointers are to be compared they must be stripped first,
because we want the result will only provide information about the address
equality, not about \invariantgroup equality.
\item Likewise, whenever a pointer is to be casted to an integer type, it must be
stripped first.
\item Whenever an integer is casted to a pointer type, the result must be
laundered before it is used, because the object stored at this address could have
changed since the last cast of the same integer to a pointer type.
\end{itemize}

These rules ensure that:
\begin{itemize}
    \item no operation can leak the pointer equality to the invariant.group model
-- as every such an operation necessarily operates on stripped pointers;
  \item every time a pointer
with a possibly unknown dynamic type is obtained, it is free of any assumptions made by the model.
\end{itemize}

\subsection{Aliasing}
\label{subsec:ModelAliasing}
% The C++ standard says that two objects with overlapping lifetimes cannot have the
% same address, unless ``one is nested within the other, or if at least one is a
% subobject of zero size and they are of different types''~\cite[\S 6.6.2.9
% Object model]{CPP}.
% However, LLVM's aliasing rules are
% weaker~\cite[Pointer Aliasing Rules]{LangRef}, e.g., both pointers are pointing
% to the same memory location, one could be a pointer to mapped memory with no write/load
% permission.
In C++, even though pointers obtained through placement new or \lstinline{std::launder} are not equivalent to the original pointer (i.e. cannot be freely substituted), they nevertheless alias the original pointer. We want to mimic this behavior in LLVM. Thus, the pointer returned by \strip or \launder aliases the intrinsic's argument:

\begin{minipage}{\linewidth}
\begin{lstlisting}[language=llvm]
store i8 42, i8* %a
%b = call i8* @llvm.launder.invariant.group(i8* %a)
; %b cannot be replaced with %a, but because they
; alias each other, we may derive %v == 42.
%v = load i8, i8* %b
\end{lstlisting}
\end{minipage}

This technique of forcing aliasing on equal-but-not-equivalent pointers allows the usual memory optimizations, such as replacing redundant loads or dead store elimination through a \launder or \strip.  The implementation simply treats \launder and \strip as bit-casts in the context of alias analysis.

We derive other properties based on aliasing of argument and returned value:
\begin{itemize}
    \item An argument of a function is considered \lstinline{nocapture} if the callee
     does not make any copies of the pointer that outlive the callee itself.
    If we know that a pointer returned from \launder or \strip is not captured,
     we can state that our intrinsic does not capture the argument.
    This is crucial, as \lstinline{nocapture} is not a valid attribute for
    return values.
    \item If the argument of a \launder or a \strip has \lstinline{nonnull} or
     \lstinline{dereferenceable} attributes, then we can also apply it to
     the returned values.
\end{itemize}

\subsection{Other Properties}
\label{subsec:Other Properties}
Since the special values \lstinline{undef} (unspecified bit-pattern) and \lstinline{null} are unable to carry any virtual information, they can be safely propagated through both of the \invariantgroup intrinsics:
\begin{itemize}
    \item \lstinline!strip(undef) == undef == launder(undef)!
    \item \lstinline!strip(null) == null == launder(null)!
\end{itemize}
In the case of \lstinline{undef}, it is useful, as it helps to remove unreachable code,
as some operations on \lstinline{undef} can be fold to the \lstinline{unreachable}
instruction, which specifies that no execution ever reaches that point in the program.
\lstinline{null} is another special value, that has an unusual treatment in LLVM, e.g. an early
exit caused by true comparison with \lstinline{null} is considered to be a cold
block (executed infrequently), which means that constant folding of \lstinline{null} is important to get the same behavior from the static profile annotation pass without any modification to specially handle the new intrinsics.

The \invariantgroup metadata also make it possible to hoist loads virtual table loads out of the loops.  This means that calling virtual function on an object in a loop will load virtual table and virtual function only once.

\subsection{Do We Need Both Intrinsics?}
\label{sec:2barriers}
With all these properties, adding calls to strip and launder related to pointer comparisons and integer-to-pointer conversions does not cause any semantic IR-level information to be lost: if any piece of information could be inferred by the optimizer about some collection of variables (e.g., that two pointers are equal), it can be inferred now about their stripped versions, no matter how many strip and launder calls have been made to obtain them in the IR.
As an example, the C++ expression \lstinline[language=C++]{ptr == std::launder(ptr)} will be optimized to true, because it is lowered to a comparison of \lstinline{strip(ptr)} with \lstinline{strip(launder(ptr))}, which are indeed equal under our rules.

One could argue that replacing uses of the \strip with \launder produces semantically correct code, although it is not strictly coherent with our model.
However, having a \strip is crucial from the optimizations point of view, as without it we lose the ability to leverage pointer equality in other optimizations.
As a toy example, \lstinline{strip(X) == strip(X)} can be folded to true, but it is not the case with \lstinline{launder(X) == launder(X)} as \launder is not pure.
\inputencoding{utf8}

\section{Outline Constructors}
\label{subsec:Outline Constructor}

External constructors pose a problem for devirtualization because after lowering to the LLVM IR it is not longer known what value is stored under a virtual pointer.
This situation appears in the following code snippet:

\begin{minipage}{\linewidth}
\begin{lstlisting}[language=C++]
struct C {
    C();
    virtual void virt_meth();
};

void foo() {
    auto *c = new C;
    // Unable to devirtualize.
    c->virt_meth();
}
\end{lstlisting}
\end{minipage}

Following Padlewski~\cite{Padlewski:2017:DL:3135932.3135947} we use ``assumption loads'', which inform the compiler about the actual dynamic type of the newly constructed object. To this end, the following three LLVM instructions are emitted after each constructor call:
\begin{enumerate}
    \item Load the virtual pointer.
    \item Compare the loaded virtual pointer with the vtable it should point to.
    \item Call \lstinline{@llvm.assume(i1 %b)} with the comparison result as a parameter.
\end{enumerate}

The intrinsic function \lstinline{@llvm.assume} takes a Boolean argument and allows the optimizer to assume that the given Boolean is true whenever the instruction is executed. In this case, it propagates equality, replacing virtual pointer loads (annotated with \lstinline{!invariant.group} metadata) with the actual value. This does not introduce any program runtime cost, as \lstinline{assume} is stripped before generating actual code, leaving comparison and load instructions trivially dead. The following code demonstrates the technique:

\begin{minipage}{\linewidth}
\begin{lstlisting}[language=llvm]
call @ctor(i8 %this_ptr)
%vtable = load %this_ptr, !invariant.group !{}
%b = icmp eq %vtable, @VTABLE_OF_C
call void @llvm.assume(i1 %b)
\end{lstlisting}
\end{minipage}

Note that, even in the case of assumption loads, the vtable load is marked with \invariantgroup to easily propagate the value.
If the constructor is inlined, the value of the comparison can easily be determined to be true, which then allows for the instructions to be optimized away.

\begin{minipage}{\linewidth}
\begin{lstlisting}[language=llvm]
; From call @ctor(i8 %this_ptr)
store @VTABLE_OF_C, %this_ptr, !invariant.group !{}
; ...
; Will be replaced with @VTABLE_OF_C.
%vtable = load %this_ptr, !invariant.group !{}
; Optimized to true.
%b = icmp eq %vtable, @VTABLE_OF_C
; assume(true) can be removed.
call void @llvm.assume(i1 %b)
\end{lstlisting}
\end{minipage}

\section{Virtual Table Definition}
\label{subsec:Virtual Table Definition}
Being able to inspect the vtable definition is necessary for full devirtualization in C++. Without it, the optimizer is only able to devirtualize vtable loads and not vfunction loads.

This is troublesome on platforms following the Itanium ABI (most of the UNIX-based platforms) as it uses the so-called ``key functions''~\cite[5.2.3 Virtual Tables]{ItaniumABI} to emit vtable in the fewest translation units possible, which is beneficial for code size and compile time.
This problem does not exist on Microsoft platforms, as the Microsoft ABI specifies that a vtable definition will be present in every translation unit that uses it.

Following the previous devirtualization models\cite{Padlewski:2017:DL:3135932.3135947}, we use the \lstinline{available_externally} linkage~\cite[Linkage]{LangRef} already present in the LLVM IR, which provides the definition of a global variable, but only for the purposes of optimizations -- the definition is stripped after optimizations leaving the variable with \lstinline{external} linkage.

\subsection{Emitting Virtual Inline Functions}
\label{subsec:Emitting Virtual Inline Functions}
Unfortunately, emitting a vtable definition without emitting all of the symbols referenced by it is not necessarily correct. This is because the Itanium ABI does not guarantee that symbols for inline virtual functions or ones marked with
\lstinline{__attribute__((visibility="hidden"))} will be exported.

Our solution is to emit vtable definitions opportunistically, viz. only when they do not reference inline virtual functions that were not emitted in that translation unit. We also implemented an option to emit all inline virtual functions with a flag \lstinline{-fforce-emit-vtable} so that all definitions of all vtables can be emitted.
However, this incurs non-trivial costs, as more functions need to be emitted and subsequently optimized. It also can increase the size of object files as some definitions might not be removed by the compiler, e.g. due to cyclic references (virtual table referencing a virtual function which in turn calls other virtual functions through the vtable).

\section{Cross Module Optimizations}
\label{sec:Cross Module Optimizations}
Whole program optimization is generally challenging, as modules compiled with a different set of optimizations could be optimized together.
This is troublesome in our case as one C++ module could be compiled with devirtualization enabled, while another module compiled from e.g. C++, C, or Swift might not use \launder and \strip where the model requires it, as can be seen in the next snippet:

\begin{minipage}{\linewidth}
\begin{lstlisting}[language=C++]
// Module 1: with devirtualization.
void with_devirt(A *a) {
    a->foo();
    A* other = unsafe_placement_new(a);
    other->foo();
}

// Module 2: without devirtualization.
A* unsafe_placement_new(A *a) {
    return new(a) B;
}
\end{lstlisting}
\end{minipage}

A naïve LTO between module 1 and module 2 could cause miscompilation, because our model assumes that \launder is emitted whenever placement new is used.

A prototype approach is to disallow linking between modules with and without devirtualization. This is not ideal as it prevents performing LTO between different languages. Another possible solution would consist of stripping intrinsics and metadata from modules with devirtualization, or introducing intrinsics in modules without it.

As future work, one can alternatively perform the said stripping lazily at function-level granularity. To this end, the \invariantgroup metadata would be equipped with a tag referring to their supported devirtualization policy, to allow future extensions and special handling of other programming languages' idiosyncracies. Moreover, functions would be annotated with a list of applicable policies. The \invariantgroup metadata and intrinsics would have no effect unless their policy is present in the policies list of the enclosing function. Otherwise, they can be safely removed. Hence, it would be possible to perform LTO across modules using \invariantgroup in different ways, e.g. a Swift module that does not need to use \launder or \strip with a C++ module using \invariantgroup virtual pointer loads. Whenever inlining or any other inter-procedural optimization is performed, the applicable policies list shall of course be set to the intersection of the policies lists of the optimization's input functions. 
\inputencoding{utf8}

\section{Benchmarks}
\label{sec:Benchmarks}
In previous work Padlewski~\cite{Padlewski:2017:DL:3135932.3135947} showed that a model that is not sound, can lead to 4 times more virtual function calls being devirtualized. 
Because the previous model's unsoundness surfaces only in hand-crafted cases that should almost never appear in real code, our model achives similar numerical results.

Our optimization was evaluated on a number of open source benchmarks using SPEC 2006, and Google internal benchmarks. In each case, the baseline was fully optimized (\lstinline{-O3}), but without FDO, LTO or post-link optimizers\cite{DBLP:journals/corr/abs-1807-06735}. 
The devirtualized build used the same setup with an additional \lstinline{-fstrict-vtable-pointers} flag.

\subsection{SPEC 2006 Benchmarks}
We did not see any difference on SPEC 2006 between optimized and devirtualized builds. This can be attributed to the fact that only half of benchmarks use C++, from which only 5 used virtual functions -- perhaps not heavily enough to show the difference in the execution time.

\subsection{Google Benchmarks}
\begin{figure*}
\centering
\includegraphics[width=0.8\textwidth]{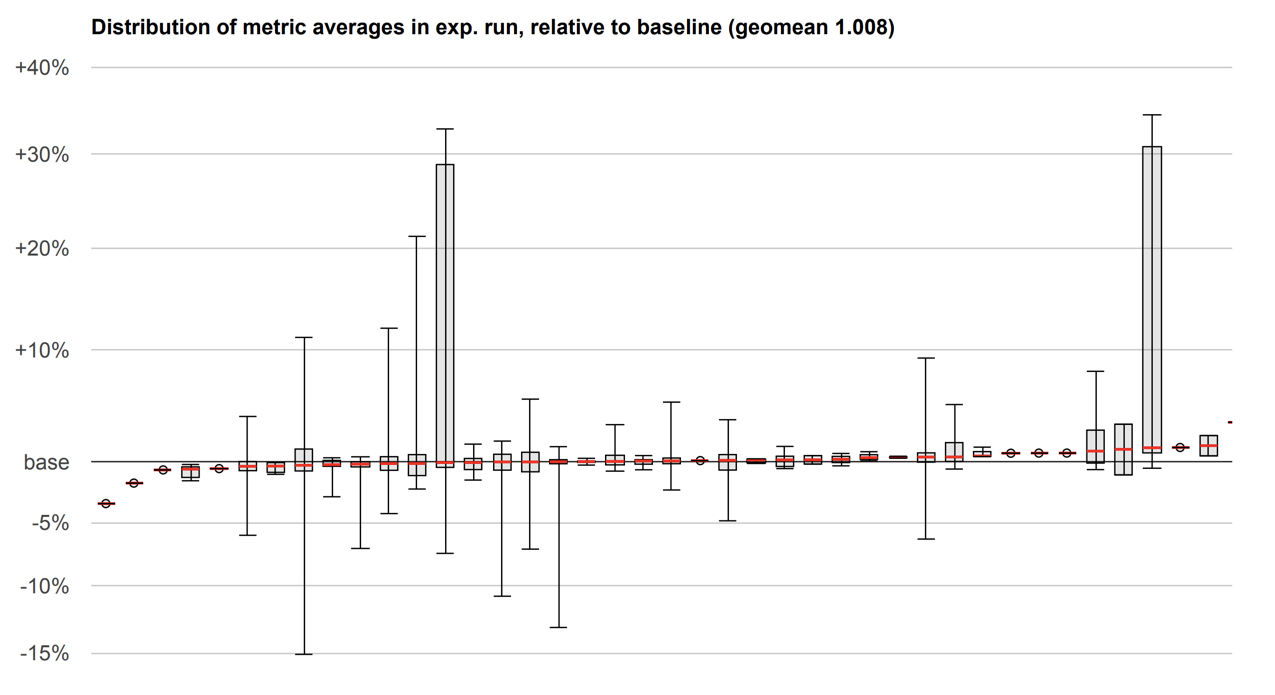}
\caption{Results of the internal sets of benchmarks. The central lines denote the medians, grey boxes correspond to the interquartile ranges and whiskers show the minimum and maximum value.}
\label{fig:geomean}
\end{figure*}

\begin{figure*}
\centering
    \includegraphics[width=0.8\textwidth]{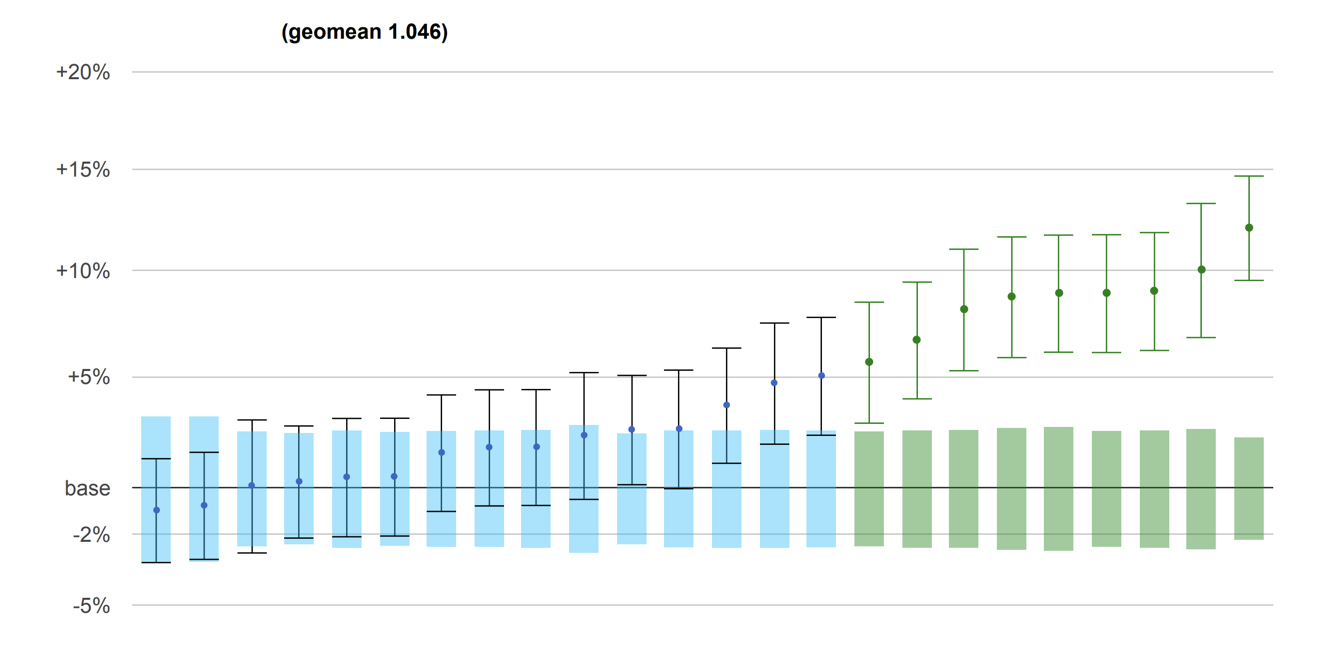}
    \caption{Average results of benchmarks within one benchmark set heavily reliant on virtual calls. 95\% confidence intervals are shown: as boxes (for base) and as whiskers (for experiment).}
    \label{fig:plaque}
\end{figure*}

We also evaluated our changes on the standard sets of internal Google benchmarks. We observed a 0.8\% improvement in the geometric mean of execution times. The results across all of the benchmarks are shown in Figure~\ref{fig:geomean}. Among these, a certain benchmark set heavily reliant on virtual calls showed a consistent 4.6\% improvement, see Figure~\ref{fig:plaque}.  The regressions seen in some benchmarks are most likely caused by the inliner making different decisions when exposed to more inlining opportunities.  This can be possibly fixed by tuning the inliner and other passes.

\begin{figure*}
\centering
    \includegraphics[width=0.8\textwidth]{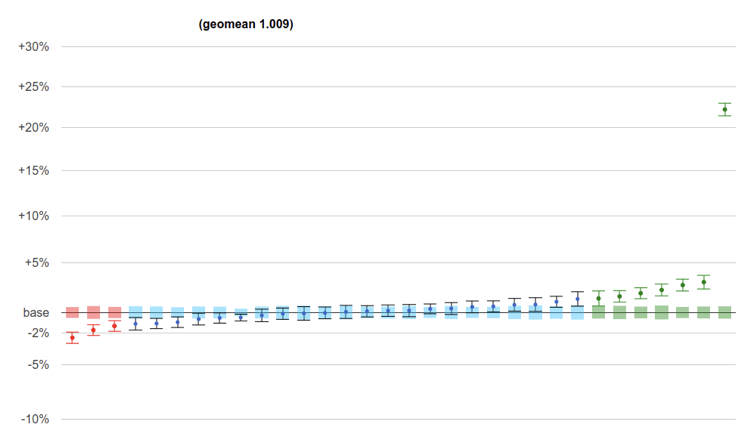}
    \caption{Average results of benchmarks within a Protocol Buffers benchmark set heavily reliant on virtual calls. 95\% confidence intervals are shown: as boxes (for base) and as whiskers (for experiment).}
    \label{fig:proto}
\end{figure*}

\begin{figure*}
\centering
    \includegraphics[width=0.8\textwidth]{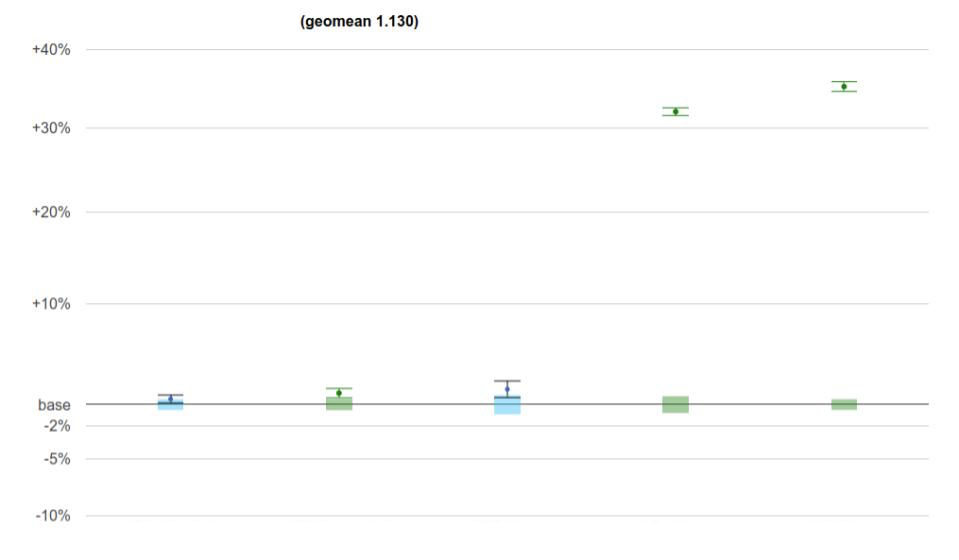}
    \caption{Average results of benchmarks within another Protocol Buffers benchmark set heavily reliant on virtual calls. 95\% confidence intervals are shown: as boxes (for base) and as whiskers (for experiment).}
    \label{fig:proto1}
\end{figure*}

Protocol Buffers is a widely-used library for serialization and deserialization of structured objects. Two sets of benchmarks for Protocol Buffers showed significant improvement, with some microbenchmarks showing over 30\% improvement. The results from one of the benchmark sets are shown in Figure~\ref{fig:proto} and Figure~\ref{fig:proto1}.

A very large internal application (Web Search) exhibits a significant 0.65\% improvement in QPS.
This strongly suggests that particularly in setups without LTO and FDO, devirtualization may bring tangible benefits.

However, in the case of LTO and FDO we haven't see improvement yet. 
We think that it mostly boils down to tunning other passes, because performing LTO or FDO on modules, that are proven to be faster should not regress them.
We do not expect to see similar improvement with LTO and FDO because they can perform subset of optimizations using more information.
It is important to note, that although some of the optimizations that we enable are theoretically possible with Whole Program Optimization, they are not feasible in practise.

We believe that better results could be seen in the languages that relay more heavily on virtual dispatch, like Java. The mentioned example of Falcon JIT~\cite{Falcon, Falcon2} achieved 10-20\% speedup relaying on unsound technique.

\section{Future Work}

Despite significant performance improvements,
our devirtualization is disabled by default in Clang, but can be enabled
with the \lstinline{-fstrict-vtable-pointers} compiler flag.
The last piece required to enable it by default is to implement the
\lstinline{supported optimizations} attribute described in
Section~\ref{sec:Cross Module Optimizations}: Cross Module Optimizations.

The described techniques rely heavily on the program's behavior being defined.
Some users are afraid of enabling optimizations that can exacerbate undefined behavior as they do not know if it is present in their code. Fortunately, a tool called Undefined Behavior Sanitizer (UBSan) detects different occurrences of undefined behavior in the runtime of the program. It would be preferable to introduce new checks to UBSan that would detect places where we falsely believe the virtual pointer did not change.

One can also model the invariance of \lstinline{const} fields in C and C++ objects.  They too can change whenever a placement new is made, so handling them necessarily requires solving the same issues as with virtual pointers. Our solution can be easily extended to this case. In the abstract model, the metadata in the fat pointer would not only store the dynamic type, but also the values of \lstinline{const} members of the object. The concrete implementation in LLVM would simply have to annotate loads and stores to \lstinline{const} variables with \invariantgroup. Moreover, \launder{}s and \strip{}s would have to be emitted for every struct type instead of only those that might have a vtable.

%\FloatBarrier

%% Bibliography
\bibliography{bibliography}{}
%\printbibliography
\bibliographystyle{plain}

\end{document}